\documentclass[aps,prl,twocolumn,showpacs,superscriptaddress,superscriptaddress]{revtex4-1}  % for review and submission
\usepackage{graphicx,color}  % needed for figures
\usepackage{dcolumn}   % needed for some tables
\usepackage{bm}        % for math
\usepackage{amssymb}   % for math

\usepackage{float}
\usepackage{amsmath}
\usepackage{epstopdf}

\begin{document}

\title{The effects of stochasticity at the single-cell level and cell size control on the population growth
}
\author{Jie Lin}
\affiliation{School of Engineering and Applied Sciences, Harvard University, Cambridge, Massachusetts 02138, USA}
\author{Ariel Amir}
\affiliation{School of Engineering and Applied Sciences, Harvard University, Cambridge, Massachusetts 02138, USA}
%\affiliation{Correspondence: arielamir@seas.harvard.edu}

%\date{\today}

\begin{abstract}
Establishing a quantitative connection between the population growth rate and the generation times of single cells is a prerequisite for understanding evolutionary dynamics of microbes.  However, existing theories fail to account for the experimentally observed correlations between mother-daughter generation times that are unavoidable when cell size is controlled for - which is essentially always the case. Here, we study population-level growth in the presence of cell size control and corroborate our theory using experimental measurements of single-cell growth rates. We derive a closed formula for the population growth rate and demonstrate that it only depends on the single-cell growth rate variability, not other sources of stochasticity. Our work provides an evolutionary rationale for the narrow growth rate distributions often observed in nature: when
single-cell growth rates are less variable but have a fixed mean, the population will exhibit an enhanced population growth rate, as long as the correlations between the mother and daughter cells' growth rates are not too strong.
\end{abstract}
\maketitle
%\dates{This manuscript was compiled on \today}
%\doi{\url{www.pnas.org/cgi/doi/10.1073/pnas.XXXXXXXXXX}}

%\begin{document}

% Optional adjustment to line up main text (after abstract) of first page with line numbers, when using both lineno and twocolumn options.
% You should only change this length when you've finalised the article contents.
%\verticaladjustment{-2pt}

%\maketitle
%\thispagestyle{firststyle}
%\ifthenelse{\boolean{shortarticle}}{\ifthenelse{\boolean{singlecolumn}}{\abscontentformatted}{\abscontent}}{}

% If your first paragraph (i.e. with the \dropcap) contains a list environment (quote, quotation, theorem, definition, enumerate, itemize...), the line after the list may have some extra indentation. If this is the case, add \parshape=0 to the end of the list environment.
\section{Introduction}
For exponentially growing populations of microbes, the population growth rate $\Lambda_p$, measuring how fast the number of individuals $N(t)$ increases with time $t$, $N(t)\sim \exp(\Lambda_p t)$, has important implications on the fitness of the population. At the single-cell level, isogenic cells also show diverse phenotypes \cite{Thattai2001,Paulsson2005,Di2007,Raj2008,Salman2012,Kiviet2014,Brenner2015,Ackermann2015}, {\it e.g.}, the generation time for a single cell to divide is heterogeneous in the population (\cite{Sandler2015,Pugatch2015,Rochman2016,Wallden2016,Taheri2015}). The mathematical relation between the heterogeneity of generation time at the single-cell level and the population-level growth rate was investigated initially by Powell (\cite{Powell1956}) with the assumption that each cell has a fluctuating generation time independent of its mother's. Powell's theory does not consider the possibility that cells in consecutive generations can be correlated, hence we refer to it as the independent generation time (IGT) model. Because the generation time sets the timing of cell division, the IGT model is also called the ``timer" model (\cite{Sauls2016}). However, recent single-cell level experiments (\cite{Wang2010,Godin2010,amir2014bending, Campos2014,Biswas2014,Cermak2016}) demonstrate that the cell volume of many bacteria in fact grows exponentially in time, e.g., {\it Escherichia coli}, {\it Caulobacter crescentus}, {\it Bacillus subtilis}. Exponentially growing cell volume is incompatible with the assumption of the absence of correlations between mother and daughter cells' generation times. In the latter case, the random noise in the generation time will lead to divergent fluctuations in the cell volume (\cite{Amir2014, amir2017cell}). Figure 1A shows direct experimental evidence for the existence of negative correlations, using data of \emph{E. coli} growth on agarose pads (\cite{Stewart2005}). Similar results have been reported previously on data taken in a microfluidic setup (\cite{Taheri2015}). These prompted us to revisit the problem, and study the effects of variability on the population growth in the presence of size control -- i.e., when mother and daughter generation times are correlated. Notably, we find that even when these correlations are weak, they have significant consequences on the population growth and it is crucial to take them into consideration -- the scenario in which there is size control, as we shall show, is fundamentally different from the ``timer" model of the IGT. As we shall highlight, these differences become especially important for the biologically relevant scenario in which growth rate fluctuations are smaller than those of the generation time (\cite{Taheri2015,Soifer2016,Kennard2016})(i.e., this appears to be the prevalent case for \emph{E. coli}, most likely the best studied organism in this context).

\begin{figure*}[bht!]
\center \includegraphics[width=1\textwidth]{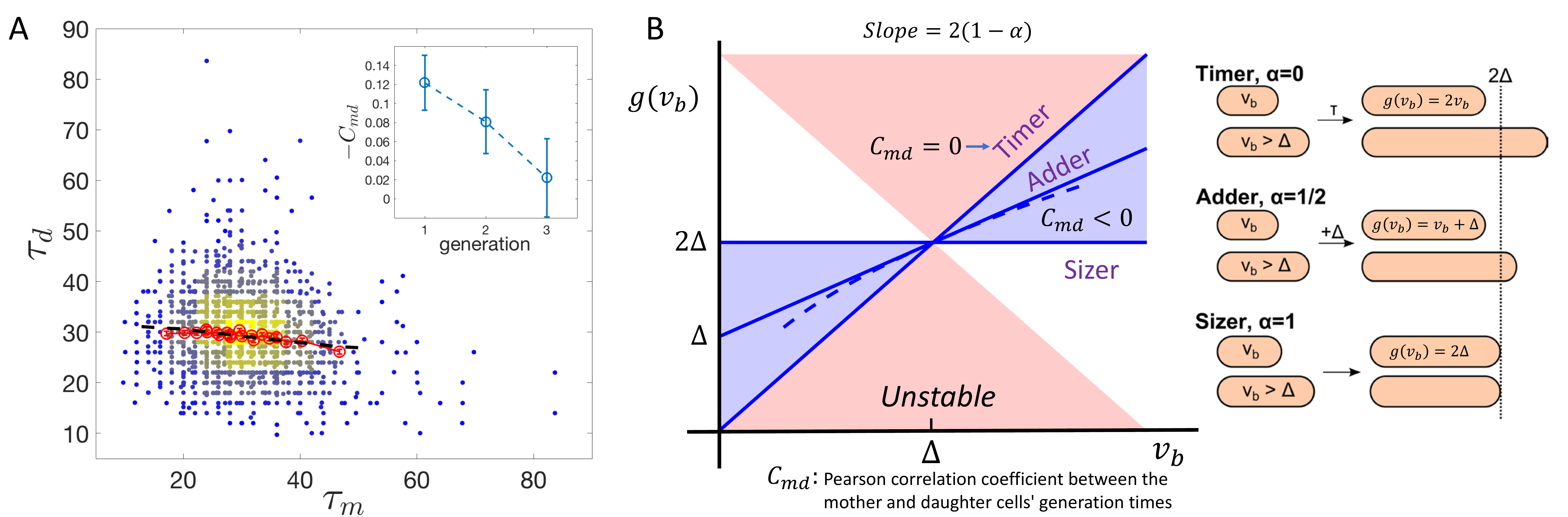}
\caption{\textbf{Phenomenological models for cell size control in bacteria.} \\
(A) Experimental measurements of the correlation between the mother ($\tau_m$) and daughter ($\tau_d$) cells' generation time.  The data used is from Ref. \cite{Stewart2005}. The color of the dots (blue to yellow) represents the local density. The red circles are the binned data and the black dashed line is the binned data of the corresponding simulations using Eqs. (\ref{size_reg}) and (\ref{tau}). Here we take $\sigma_{\lambda}=0.19$, $\sigma_{\xi}=0.12$, and the resulting CV of generation time is about $0.3$, similar to the experimental data. Inset: the (negative) Pearson correlation coefficient between mother and daughter generation times plotted as function of the generation difference between the progenitor and descendent cells. The first point corresponds to the mother-daughter correlation.\\
(B) The size at division, $g(v_b)$, depends on the size at birth (x-axis), and the parameter $\alpha$ is related to the slope of the function, and can continuously interpolate across different models. $\Delta$ is the average cell size at birth. Here we show three examples. As $\alpha$ increases from 0 to 1, the negative correlations between the generation time of mother and daughter cells become stronger. Red regions are unstable in terms of cell size regulation. Figure adapted from Ref. \cite{amir2017cell}. The dashed line corresponds to the approximate model of Eq. (\ref{approxeq}) (STAR Methods) for $\alpha=1/2$.\\}
\label{sizeregulation}
\end{figure*}

The structure of the paper is as follows. We first provide the necessary background for our implementation of the size control (using the discrete Langevin approach, recently introduced in this context (\cite{Amir2014})) and the various sources of stochasticity in the problem: time-additive noise to the generation time, size-additive noise to the division size and growth rate variability which will turn out to be the most consequential noise term. Since this is a rather formidable problem mathematically, we first provide results for the limit of vanishing growth rate fluctuations, which will allow us to gain intuition and will elucidate why the size control plays such a significant role. We are able to show that as long as some form of size control exists, in this case the population growth rate is equal to the single-cell growth rate. Next, we show our central result: irrespective of the strength of the size control, in the presence of cell size regulation the population growth rate is only determined by the variability of single-cell growth rates and its correlation time, and is independent of the strength of size control or other sources of stochasticity in the generation time. We find that as long as the growth rate is not too strongly correlated between mother and daughter cells, the population growth rate is \emph{decreased} due to the growth rate variability. In the case of uncorrelated growth rates, we are able to provide a closed analytic expression for the population growth rate, which we corroborate with numerical simulations. Our analysis of \emph{E. coli} growth data shows weak growth rate correlations, allowing us to provide a detailed comparison between our analytic results and the experiments.
{ Throughout this work, we assume that cells are kept in a constant environment, and neglect effects of cell crowding and cell death -- all cells divide and give rise to two offspring, generating an exponentially growing lineage tree. }

%For the IGT model, the generation time of every node in this expanding tree is independently drawn from some distribution, yet in our case the negative correlations in mother and daughter generation times will create a complex pattern of correlations between relatives on this tree, and will also bias the statistics (since cells with shorter generation time will have an advantage in proliferating). In fact, our analysis reveals intricate differences and connections between generation time distributions measured over single lineages and population trees, which may be viewed as the underlying reason for the failure of the IGT model in providing an adequate description of realistic scenarios}.

\subsection{The size regulation model}
To make sure cell size is regulated, cells must adopt some regulation strategy (\cite{Osella2014,Robert2014}). We assume cells attempt to divide symmetrically at the division size, $v_d=g(v_b)$ and $v_b$ is the cell size at birth. For the sizer model where there is a critical size for cells to divide (\cite{Cooper1991,Koch2001}), $g(v_b)=$ const, while the timer model corresponds $g(v_b)=2v_b$ since the single-cell growth is exponential. Recent experiments (\cite{Campos2014,Taheri2015, Deforet2015,Soifer2016,Kennard2016}) observe that cell divisions of many microorganisms are in fact regulated by the adder model (\cite{Amir2014,Ghusinga2016,Kennard2016}), where cells attempt to add a constant volume $\Delta$ to its birth size, $g(v_b)=v_b+\Delta$. It has also been argued that the adder model is intimately related to the regulation of genome replication (\cite{Amir2014, Ho2015, zheng2016interrogating, amir2017cell}).

In the following, we choose a simple regulation model which can unify the three strategies as
\begin{equation}
v_d=2\alpha \Delta+2(1-\alpha) v_b,
\label{size_reg}
\end{equation}
where $\Delta$ is a constant and $\alpha$ is the regulation parameter. It follows directly that $\alpha=0,\frac{1}{2}, 1$ correspond respectively to the timer, adder, and sizer model (\cite{Amir2014}). The above three models are illustrated in Figure \ref{sizeregulation}B. Given the division size and exponential growth at the single-cell level, we can find the corresponding generation time, for which we consider two sources of noises in the main text, the growth rate fluctuations and the time-additive noise $\xi$, such that
\begin{equation}
\tau=\frac{1}{\lambda}\ln \left( \frac{v_d}{v_b}\right) +\xi ,\label{tau}
\end{equation}
where the single-cell growth rate $\lambda$ is assumed to be distributed normally with mean $\lambda_0$ and variance $\sigma_{\lambda}^2$. Similarly, the time-additive noise $\xi$ is assumed to satisfy a normal distribution with zero mean and variance $\sigma_{\xi}^2$. { Later we shall show the robustness of these results to other forms of noise, namely, size-additive noise, non-Gaussian time-additive noise as well as a non-Gaussian growth rate distribution (STAR Methods).}

\begin{figure*}[bht!]
\center    \includegraphics[width=0.85\textwidth]{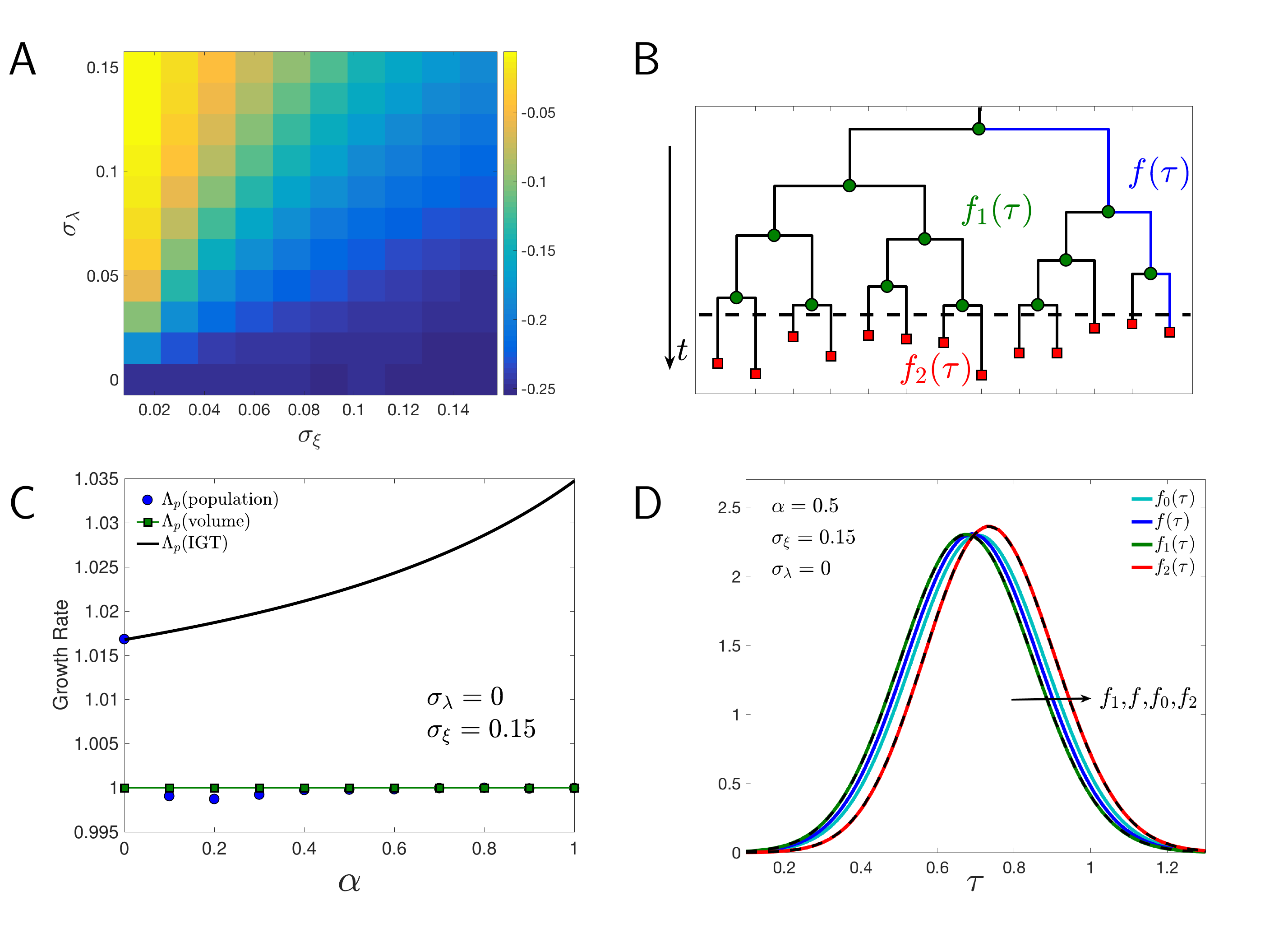}
\caption{\textbf{Differences between the statistics on single lineages and on trees and the breakdown of the Independent Generation Time  (IGT) model.}
\\(A) Pearson correlation coefficient between the mother and daughter generation times, $C_{md}$, as function of the time-additive noise standard deviation $\sigma_{\xi}$ and growth rate variability standard deviation $\sigma_{\lambda}$. We simulated the regulation model for $\alpha=1/2$. $C_{md}\approx -0.25$ for $\sigma_{\xi}>0$, $\sigma_{\lambda}=0$. These finite correlations between mother and daughter cells' generation times invalidate the assumptions behind the IGT model\\
(B) A lineage tree starting from one cell. Division events are labeled as green circles and red squares. To define the generation time distribution, one can track along a single lineage, see the blue lines, corresponding to the generation time distribution $f(\tau)$. We may also take a snapshot of the population, shown as the dashed line, and record all the current and past cells on the tree. Those cells which have already divided live on the branch nodes, so we denote them as branch cells (green circles), and the corresponding generation time distribution as $f_1(\tau)$. Those cells which have not divided yet live on the leaf nodes (and they will divide later, as marked by the red squares), so we denote them as leaf cells (red squares), and their generation time distribution is denoted $f_2(\tau)$. \\
(C) The independent generation time (IGT) model breaks down under cell size regulation ($\alpha>0$). Numerical simulations of the regulation model, with $\sigma_{\xi}=0.15$, $\sigma_{\lambda}=0$ (i.e., no growth rate fluctuations), and $\lambda_0=1$. The initial number of cells is $N_0=1000$. The relative age of initial cells, the time after the cell's birth divided by the generation time, is uniformly distributed between $0$ and $1$. Data is collected at time $t=10$. The same numerical protocol is used in the following numerical simulations. The population growth rate is independent of $\alpha$ for $\alpha>0$ with an error bar smaller than the marker size. The systematic small deviations, especially for weak size regulation (small $\alpha)$, is presumably due to the finite simulation time. The volume growth rate precisely equals the single-cell growth rate $\lambda_0=1$, as we prove in the main text. The black solid line is the IGT model's prediction, Eq. (\ref{IGT_formula}).\\
(D) Generation time distributions, respectively defined for (i) all cells on the tree $f_0$, (ii) cells along lineages $f$, (iii) only branch cells $f_1$, (iv) only leaf cells $f_2$, see Fig. \ref{lineages_vs_trees}B. The solids lines are extracted from numerical simulations, and the order of the four distributions is indicated by the arrow. Theoretical predictions of $f_1$ and $f_2$, according to Eq. (\ref{f1}), (\ref{f2}) are shown as the dashed lines.}\label{lineages_vs_trees}
\end{figure*}

When cell size is regulated, the correlation coefficient between the mother-daughter generation time becomes nonzero (\cite{Taheri2015}), in contrast to the central assumption of the IGT model. Despite the simplicity of the regulation model, it is challenging to find the exact stationary distribution of birth volumes and generation times (\cite{marantan2016stochastic}). To make the theoretical analysis feasible, we consider an approximate version of the regulation model (\cite{Amir2014}) (Eq. (\ref{approxeq}), STAR Methods).
%
%\begin{equation}
%v_d=2\Delta ^{\alpha} v_b^{1-\alpha}. \label{approxeq}
%\end{equation}
%
%It is straightforward to verify that for any $\alpha$ this agrees with Eq. (\ref{size_reg}) to first order when expanded around the typical birth-size. Furthermore, the coefficient of variation (CV, the standard deviation divided by the mean) of cell birth sizes are often reported to be around $0.1$ \cite{Campos2014}, indicating that the noise is relatively small and that the first order expansions makes for an excellent approximation, {\blue see the dashed line in Figure 1B}. Thus, the approximate model provides a very good realization of the original model. The corresponding generation time becomes
%\begin{equation}
%\tau=\frac{\ln 2}{\lambda}-\frac{\alpha}{\lambda}\ln \left(\frac{v_b}{\Delta}\right)+\xi\label{tauapprox}.
%\end{equation}
We are able to calculate the correlation of mother-daughter generation time in the limit $\sigma_{\lambda}=0$ for the approximate model
\begin{equation}
C_{md}=\frac{\langle\tau_m\tau_d\rangle -\langle\tau_m\rangle \langle\tau_d\rangle}{\sigma_{\tau}^2}= -\frac{\alpha}{2},\label{cmd}
\end{equation}
where $\tau_m$($\tau_d$) is the generation time of mother (daughter) cells (STAR Methods). For the timer case ($\alpha=0$), the mother-daughter correlation is zero and the IGT model is valid as expected. However, for $\alpha>0$, the mother-daughter generation time is negatively correlated. In the more general case of finite growth rate fluctuations, where analytical results are not attainable, we note that the negative correlations tend to be suppressed, see Figure \ref{lineages_vs_trees}A, showing the dependence of the correlation coefficient on both noise terms. In the opposite limit $\sigma_{\xi}=0$, $\sigma_{\lambda}>0$ (which, as mentioned above, appears to be less relevant biologically) the cell size converges to approximately $\Delta$ immediately and the generation time is determined by the growth rate for any $\alpha$, in which case the mother-daughter correlation becomes zero and the IGT model is recovered as well. For the ``adder" scenario, $\alpha=1/2$, we have $-0.25<C_{md}<0$. This is consistent with the analysis of the experimental data of {\it E. coli} (\cite{Stewart2005}) shown in Figure \ref{sizeregulation}A, and corroborated with numerical simulations based on Eqs. (\ref{size_reg}) and (\ref{tau}).
%Notations of the size regulation model are summarized in Table. S1.
\section{Results}

\subsection{A simple case: $\sigma_{\lambda}=0$}
First, we focus on the size regulation model in a particular limit that will be instructive to elucidate the discrepancy with the IGT model. Consider a simpler situation without growth rate fluctuations, such that all cells grow at the same rate $\lambda_0$. One can track a single lineage for many generations, shown as the blue lines in the lineage tree in Figure \ref{lineages_vs_trees}B. As the generation number increases, the distributions of $\tau$ and $v_b$ converge to the stationary distribution. In this case, for small noise the mean cell size at birth is approximately $\Delta$, and the variance of $\ln(v_b/\Delta)$ is approximately $\sigma_{v}^2=\lambda_0^2\sigma_{\xi}^2/(2\alpha-\alpha^2)$ (\cite{Amir2014}). Thus, the cell size distribution will remain bounded as long as $0<\alpha<2$.

In a seminal paper, Powell has worked out an elegant relation between the generation time distribution along lineages, $f(\tau)$, and the population growth within the IGT model, which is still widely used (\cite{Powell1956}):

\begin{equation}
2\int_{0}^{\infty} f(\tau) e^{-\Lambda_p\tau} d\tau=1 .\label{growthrateMF}
\end{equation}
(for completeness, we provide the derivation of this relation in the STAR Methods).

Within the IGT, the distribution of generation times along a single lineage is identical to that of the entire population tree. We will now show that this is no longer the case when size control induces correlations across generations, which raises the question: Does a similar relation to Eq. (\ref{growthrateMF}) hold, and if so, what is the correct distribution to use? We find that using the generation time distribution along lineages, $f(\tau)$, to infer the population growth rate using Eq. (\ref{growthrateMF}) is incorrect, and leads to qualitatively erroneous results.

We first consider the growth of the total cell volume, with a volume growth rate, $V(t)=\sum_{i=1}^{N(t)} v_i(t)\sim \exp(\Lambda_v t)$. In the case without single-cell growth rate fluctuations (but in the presence of generation time variability), { we have at any time point other than cell division events:}
\begin{equation}
\frac{dV(t)}{dt}=\sum_{i=1}^{N(t)} \frac{d v_{i}(t)}{dt}=\lambda_0 V, \label{vgrowth}
\end{equation}
so $V(t)=V_0\exp(\lambda_0 t)$, where $V_0$ is the total volume at time $t=0$. Since the total volume of cells is continuous also during cell division events, { we conclude that cell volume increases exponentially with rate $\lambda_0$. }

The average cell size of the instantaneous population at time $t$ thus changes as $\bar{v}\sim \exp((\lambda_0-\Lambda_p)t)$. Since the cell size is regulated, we immediately obtain that $\Lambda_p=\lambda_0$, regardless of the time-additive noise $\sigma_{\xi}$, and the regulation parameter $\alpha$.
On the other hand, the generation time distribution along lineages $f(\tau)$ is well approximated by a normal distribution with mean $\tau_0=\ln(2)/\lambda_0$ and variance $\sigma_{\tau}^2=2\sigma_{\xi}^2/(2-\alpha)$ (\cite{Amir2014}). Thus we can calculate $\Lambda_p$ as predicted by Eq. (\ref{growthrateMF}) ,
\begin{equation}
\Lambda_{p}(IGT)=\lambda_0 \frac{2}{1+\sqrt{1-\frac{2\sigma_{\tau}^2\lambda_0^2}{\ln2 }}}.\label{IGT_formula}
\end{equation}
It is easy to show that $\Lambda_p(IGT)>\lambda_0$ if $\sigma_{\tau}>0$, contradicting the exact result $\Lambda_p=\lambda_0$. We thus exemplify the shortcoming of the IGT in the presence of size control.

We simulated the size regulation model, starting from $N_0=1000$ cells, and computed the population and volume growth rate in the exponential phase. Indeed, we find that the volume growth rate $\Lambda_v$ precisely equals $\lambda_0$, and that the population growth rate $\Lambda_p$ matches the volume growth rate for $\alpha>0$, as shown in Figure \ref{lineages_vs_trees}C. The IGT model's prediction exceeds the numerical results, except for the case of no size control, $\alpha=0$, when it is exact. { Since the volume growth rate is equal to the population growth rate in the presence of cell size regulation, in principle one could use either of them to infer the population growth rate. However, the fact that $V(t)$ increases continuously makes it a favorable candidate to be used both in our numerical results as well as in our analysis of experimental data, shown later. Using it on a limited amount of data will result in far more accurate results.}

\subsection{Single lineages and tree statistics are distinct} Tracking a single lineage is not the only way to define the generation time distribution. Given a lineage tree, one can take a ``snapshot" at any time, marked as the dashed line in Figure \ref{lineages_vs_trees}B. To be precise, we need to differentiate two kinds of cells on the tree: (i) cells in the present snapshot, (ii) cells that have divided prior to the present snapshot. Since they respectively live on the leaves and branches of the tree, we denote them as leaf cells and branch cells respectively, marked as red squares and green circles in Figure \ref{lineages_vs_trees}B. Besides the distribution along single lineages ($f(\tau)$), we define additional three generation time distributions for the branch cells ($f_1(\tau)$), the leaf cells ($f_2(\tau)$), and all cells on the tree ($f_0(\tau)$), respectively. It is only when there are no mother-daughter correlations that $f(\tau)=f_0(\tau)$, otherwise, these four distributions are all distinct as shown in Figure \ref{lineages_vs_trees}D. We will show that the seemingly innocuous differences between the different distributions will have major consequences for the role of variability on the population growth. Compared with $f_0(\tau)$, the leaf cell distribution $f_2(\tau)$ is biased towards cells with longer generation times because they have a larger chance to be observed in a snapshot. Conversely, $f_1(\tau)$ is biased towards those cells with shorter generation time (\cite{Powell1956,Wakamoto2012,Hashimoto2016}). Mathematically, $f_1(\tau)$, $f_2(\tau)$ are related to $f_0(\tau)$ by
\begin{align}
f_1(\tau)&=2e^{-\Lambda_p \tau}f_0(\tau),\label{f1}\\
f_2(\tau)&=2(1-e^{-\Lambda_p \tau}) f_0(\tau). \label{f2}
\end{align}
We provide a detailed derivation in the STAR Methods. Eqs. (\ref{f1}),(\ref{f2}) are verified numerically for the adder model in Figure \ref{lineages_vs_trees}D. Moreover, we can show that in the case where there is a finite mother-daughter correlation, one can still make use of the IGT model's result, and the only difference is to replace $f(\tau)$ by $f_0(\tau)$,
\begin{equation}
2\int_{0}^{\infty} f_0(\tau) e^{-\Lambda_p\tau} d\tau=1 .\label{growthratef0}
\end{equation}
The detailed derivation is in the STAR Methods and we discuss a numerical test of this prediction in the next section, where we also work out the consequences of these intricate relations.

\begin{figure*}[bht!]
\center   \includegraphics[width=1\textwidth]{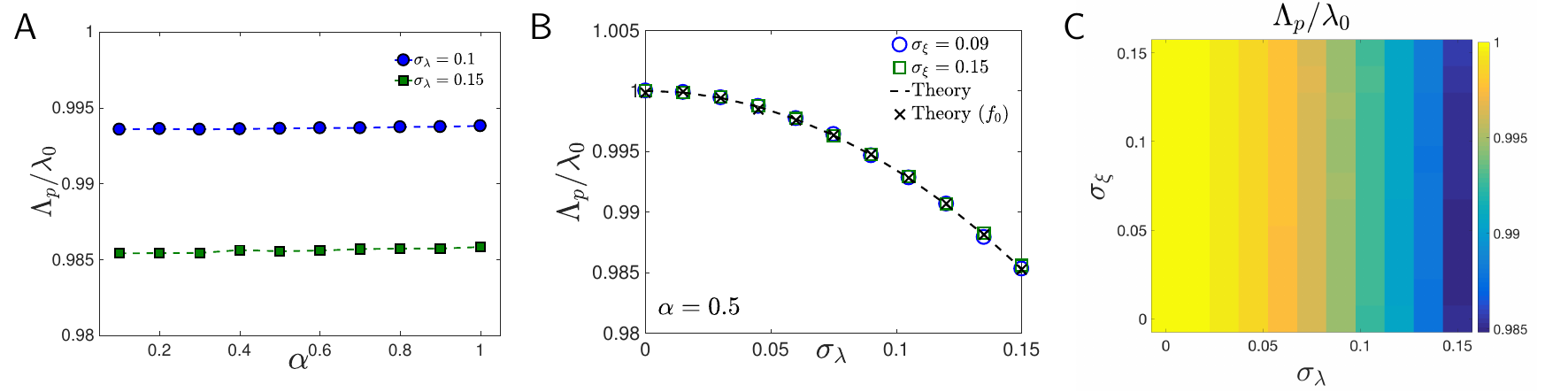}
\caption{\textbf{Population growth with variable single-cell growth rates.} \\
(A) $\Lambda_p/\lambda_0$ \emph{vs.} $\alpha$, for $\sigma_{\xi}=0.15$, $\lambda_0=1$, and $\sigma_{\lambda}$ as indicated. We do not see systematic dependence of $\Lambda_p$ on $\alpha$, and the differences between $\Lambda_p$ at different $\alpha$ that we find in our simulations are of the order $10^{-4}$. \\
(B) $\Lambda_p/\lambda_0$ \emph{vs.} $\sigma_{\lambda}$ for $\alpha=1/2$, $\lambda_0=1$, and $\sigma_{\xi}>0$. Data with different $\sigma_{\xi}$ collapse on the same curve, implying that $\Lambda_p$ is independent of $\sigma_{\xi}$. The theoretical prediction Eq. (\ref{growthrate1}) (dashed line) matches the numerical results very well. We also show the theoretical prediction from the tree distribution $f_0$ according to the modified IGT formula, Eq. (\ref{growthratef0})\\
(C) Same plot as (B) where a range of time-additive noises $\sigma_{\xi}$ are shown, and we find that the population growth rate only depends on the single-cell growth rate variability $\sigma_{\lambda}$.}
%(D) The relative reduction of population doubling time $T_d$ agaist the mean generation $\langle \tau\rangle$. The colorful surface plot is the result from the regulation model with $\alpha=1/2$. The dashed surface is the IGT model's prediction using Eq. (\ref{growthrateMF}). They coincide only if $\sigma_{\xi}=0$, when the mother-daughter correlation vanishes. Otherwise, the correction predicted by the IGT model is much larger than the regulation model's prediction. \\
%(D) Single-cell growth rates {\it v.s.} the generation times of a population, obtained from our simulations. The Pearson correlation coefficient is about $-0.38$. The red circles are the binned data and the dashed line is an empirical fitting, $\tau\approx 0.74/\lambda$.
%(E) Population growth rate $\Lambda_p$ {\it v.s.} $\sigma_{\lambda}$, in the presence of size-additive noise. $\sigma_{\eta}$ is the standard deviation of the size-additive noise, which is shown to have no effect on the population growth rate. The dashed line is the theoretical prediction of Eq. (\ref{growthrate1}). Here, we fix the standard deviation of the time-additive noise to be $\sigma_{\xi}=0.05$. For (A), (B), (E), the error bars are smaller than the marker sizes.}
\label{variable_growth_rate}
\end{figure*}

\subsection{Variable single-cell growth rates: $\sigma_{\lambda}>0$}
We now turn to the general case with a finite variability in the single-cell growth rates. Motivated by the biologically relevant scenarios (as we shall elaborate on in the next section), we first focus on the case where growth rates are uncorrelated between mother and daughter cells. The case of correlated growth rates will be considered later.

In general, the population growth rate, $\Lambda_p$, is a function of the three variables $\alpha$, $\sigma_\xi$, and $\sigma_\lambda$, and we have shown that when $\sigma_{\lambda}=0$, $\Lambda_p(\alpha,\sigma_{\xi},0)=\lambda_0$ for any $\sigma_{\xi}$, and $\alpha>0$. Notably, we find numerically that $\Lambda_p$ is again independent of $\alpha$ for a finite $\sigma_{\lambda}$, as shown in Figure \ref{variable_growth_rate}A. In the same way, we find that $\Lambda_p$ is independent of $\sigma_{\xi}$ as well, shown in Figure \ref{variable_growth_rate}B,C. Thus, $\Lambda_p$ appears to be a function of $\sigma_{\lambda}$, irrespective of $\sigma_{\xi}$ and $\alpha$. The fact that $\Lambda_p$ is independent of $\sigma_{\xi}$ allows us to obtain the general expression of $\Lambda_p$, since $\Lambda_p(\alpha,\sigma_{\xi},\sigma_{\lambda})=\Lambda_p(\alpha,0,\sigma_{\lambda})$. As mentioned previously, the mother-daughter correlations vanish in the limit $\sigma_{\xi}\rightarrow 0$, wherein the generation time is merely determined by the growth rate, $\tau=\ln(2)/\lambda$. Since the growth rates of mother and daughter cells are assumed to be independent, mother and daughter generation times will be uncorrelated in this limit, hence $f_0(\tau)=f(\tau)$. We can therefore calculate the theoretical value of $\Lambda_p$ using Eq. (\ref{growthratef0}) and $\tau=\ln(2)/\lambda$, namely $2\int_{0}^{\infty}\rho(\lambda)\exp(-\ln(2)\Lambda_p/\lambda)=1$ where $\rho(\lambda)$ is the distribution of single-cell growth rates. For small $\sigma_{\lambda}$, we can compute the analytic expression of the population growth rate using the saddle point approximation (STAR Methods)
\begin{equation}
\Lambda_p(\sigma_{\lambda})=\lambda_0\{1-\left(1-\frac{\ln2}{2}\right)\left(\frac{\sigma_{\lambda}}{\lambda_0}\right)^2\},\label{growthrate1}
\end{equation}
verified in Figure \ref{variable_growth_rate}B. We also test the more fundamental result, Eq. (\ref{growthratef0}) in Figure \ref{variable_growth_rate}B and illustrate the erroneous prediction of the IGT model's Eq. (\ref{growthrateMF}) in Figure S1F. Finite growth rate fluctuations thus tend to {\it decrease} the population growth rate given a fixed mean value. Experimentally the coefficient of variation (CV) of single-cell growth rate, $\sigma_{\lambda}/\lambda_0$ which varies from $6\%$ to $20\%$, has been reported for {\it E.coli} in different growth conditions (\cite{Taheri2015,Cermak2016, Wallden2016,Kennard2016}), significantly smaller than the CV of generation times ($20\%-40\%$) (\cite{Taheri2015, Hashimoto2016}). The above values of CV for the growth-rate variability can generate a $0.2\%$ to $3\%$ growth deficit according to our predictions, such an effect is expected to be significant for evolutionary dynamics, for which “strong selection” is defined as the regime where mutation effects are larger than the inverse of the population size (\cite{Orr2002,Nowak2006}). Our results may explain the origin of the small growth rate variability as it improves the overall fitness of the population, while no such evolutionary pressure is exerted on the generation time distribution itself, in consistence with the broader distributions observed experimentally.

An intuitive way to understand the reduction of population growth rate is to realize that the growth rates of the leaf cells (an instantaneous snapshot) are negatively correlated with the leaf cells' generation times $\tau$. Since $e^{\lambda \tau} \approx 2$, one would naively expect $\tau \approx \log(2)/\lambda$ for the binned correlations, consistent with our simulations (see Figure S1G). Those cells with smaller growth rates have a larger generation time and therefore a larger chance to be observed in a snapshot. Because the leaf cells are biased towards cells with smaller growth rates, the mean growth rate of leaf cells decreases, and so does the population growth rate.
Mathematically, we can re-write Eq. (\ref{vgrowth}) as:
\begin{equation} \frac{dV}{dt} = \sum_{\textrm{all \emph{leaf} cells}} \lambda_i v_i  ,\end{equation}
and since the growth rates are drawn for each cell independently of its volume we do not expect $\lambda_i$ and $v_i$ to be strongly correlated (numerically, we indeed find a correlation coefficient close to zero), suggesting that when taking the average we may replace $v_i$ with the typical cell size. This would suggest that the growth rate may be approximated by the average of the growth rate over all leaf cells  (i.e., cells growing at the time of observation), which as argued above would be biased towards favoring the smaller growth-rates.
%We will now analyze this effect quantitatively.
A recent study on the particular case of the sizer model led to similar conclusions (\cite{Olivier2016}). We show here that this extends to the biologically relevant case of the adder model, and in fact, that the results are independent of the strength of size control.

%However, the relative difference between the two variables is significantly more moderate than the IGT model's predictions, a result which can be tested on single-cell measurements.
\section{Robustness of the results to the form of noise distributions}
 One may wonder whether the assumption of normality of the time-additive noise and the growth-rate fluctuations provides a good approximation for realistic scenarios. As we have shown, the time-additive noise to the generation times has no effect on the population growth. Effects of size-additive noises on the division sizes are discussed in the STAR Methods, which turn out to be irrelevant as well (Figure S1). Note that the size-additive noises results in non-Gaussian generation time distributions, with an approximately exponentially tail, similar to that reported in Ref. \cite{Pugatch2015} (Figure S2). This suggests that our results should be robust to the precise nature of the noise affecting the generation time. Indeed, we show that replacing the Gaussian time-additive noise by a skewed Gamma distributed time-additive noise has no effect on the population growth (STAR Methods, Figure S1).

\begin{figure*}[bht!]
\center  \includegraphics[width=1\textwidth]{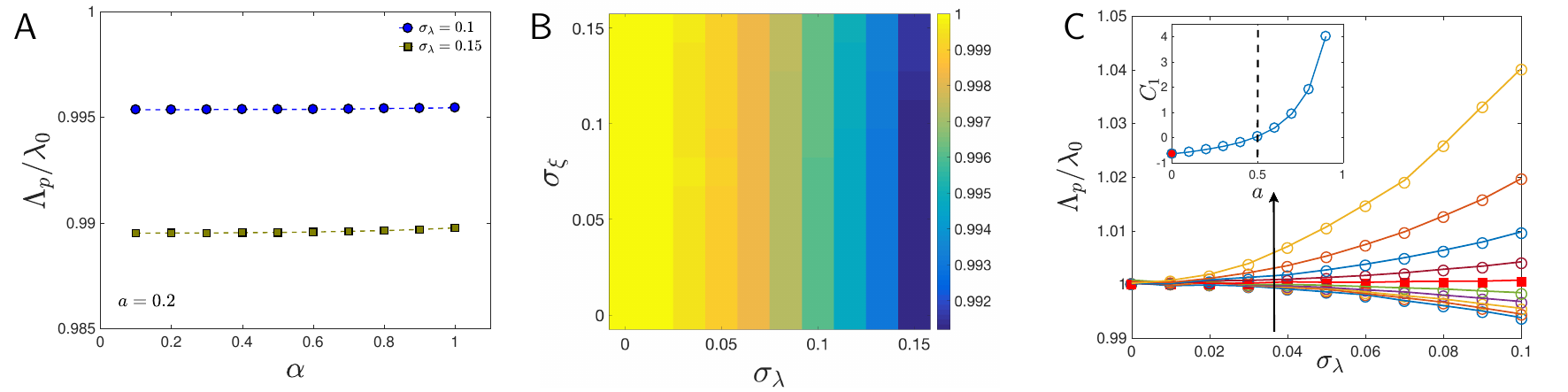}
\caption{\textbf{Dependence of the population growth rate on different noises for correlated growth rates.}\\
(A) $\Lambda_p/\lambda_0$ \emph{vs.} the size control parameter $\alpha$, for $\sigma_{\xi}=0.15$, $\lambda_0=1$, $\sigma_{\lambda}$ and $a=0.2$ as indicated.\\
(B) Two dimensional dependence of $\Lambda_p$ on the time-additive noise $\sigma_{\xi}$ and the growth rate variability $\sigma_{\lambda}$. $a=0.2$. One can see that the population growth rate is only dependent on single-cell growth rate variability $\sigma_{\lambda}$.\\
%(C) Similar as
 (B), with the time-additive noise replaced by the size-additive noise $\sigma_{\eta}$.\\
(C) Population growth rate $\Lambda_p/\lambda_0$ \emph{vs.} $\sigma_{\lambda}$ for a range of $a$ from $0$ to $0.9$ with a fixed step $0.1$. The direction of increasing $a$ is indicated by the arrow. $\sigma_{\xi}=0.1$. The solid lines are the predictions from the tree distribution Eq. (\ref{growthratef0}). The threshold case $a=0.5$ is shown in red, and the inset shows the dependence of the fitting parameter $C_1$ on $a$ from Eq. (\ref{growthratea}).
}\label{correlated_growth_rate}
\end{figure*}
\begin{table*}
\caption{Summary of the main theoretical and numerical results of this work. $\alpha$ and $a$ are respectively the cell size control parameter and growth rate correlation coefficient.}
\begin{tabular}{ l | c   c   c   c}
\hline
 &   $\alpha=0$ \quad & $\alpha>0$, $a=0$ \quad \quad  & $\alpha>0$, $a<1/2$  \quad\quad & $\alpha>0$, $a>1/2$  \\ [3pt] \hline
 Size control &  No  & Yes  &  Yes   & Yes \\ [3pt] \hline
N(t) and V(t) grow exponentially \\ at the same rate  & No  & Yes  & Yes   & Yes  \\[3pt] \hline
Mother-Daughter Generation \\ times independent & Yes  & No  & No   & No  \\[3pt] \hline
Mother-Daughter Growth rates\\ independent  & Irrelevant  & Yes  & No   & No  \\[3pt] \hline
Effect of  cell cycle variability \\ on population growth for fixed \\growth rate distribution & Positive  & None   & None  & None\\[3pt] \hline
Effect of growth rate variability \\ on population growth  & Irrelevant  & Negative    & Negative & Positive  \\[3pt] \hline
Tree statistics different than \\ single lineage statistics  & No  & Yes   & Yes &Yes \\[3pt] \hline
Equation determining population growth & Eq. (\ref{growthrateMF})  & Eq. (\ref{growthratef0}), Eq. (\ref{growthrate1})  & Eq. (\ref{growthratef0})  & Eq. (\ref{growthratef0})  \\[3pt] \hline
\end{tabular}\label{table1}
\end{table*}
Although measured distribution of growth rates are not precisely Gaussian, they are well approximated by a Gaussian distribution (\cite{Wang2010,Osella2014,Taheri2015,Wallden2016}). Indeed, as we shall show in the experimental test later, using the distribution of growth rates directly inferred from the experimental data of Stewart et al. (\cite{Stewart2005}) (without assuming Gaussian statistics) leads only to minor differences in the results, showing that the assumption of Gaussian statistics is a reasonable one in this case. For completeness, in the STAR Methods we also report the numerical results for a log-normal distribution of growth rates, demonstrating that our conclusions regarding the role of variability on the population growth rate remain intact (Figure S1). { We note that Eq. (\ref{growthrate1}) is correct also for the case of log-normal distributed growth rates, if $\lambda_0$ is replaced by the \emph{mean} growth rate (STAR Methods).}
In order to fully specify the model, we need to define the correlations of growth rates across generations. So far, we have assumed that the growth rate is constant throughout each cell cycle, and independent of the growth rate of previous generations. This assumption will provide an excellent approximation for \emph{E. coli} growth, where correlations between mother and daughter growth rates are small, as we elaborate on in the next section. For completeness, in the next section we also relax this assumption and study the intriguing effects of strong correlations in growth rates across generations.
%Finally, we note that in various cases the relative growth rate fluctuations are smaller than the relative generation time fluctuations. Our results provide a potential explanation of this fact, since only the former noise term will affect the population growth.
%We emphasize this point here, as this fact has led us to put more emphasis on this regime (i.e., $\sigma_\lambda < \sigma_\xi$) which most of our simulations have centered on.

\section{Effects of growth rate correlations}
%So far we assumed that there is no correlation between the growth rates of mother and daughter cells.
In this section, we generalize the model to include a finite positive correlation in growth rates. We assume the daughter cell's growth rate $\lambda_d$ is dependent on its mother's growth rate $\lambda_m$, and also subject to random noise,
\begin{equation}
\lambda_d=a\lambda_m+b+\epsilon\label{a}
\end{equation}
here $a$ is a number between $0$ and $1$, and $\epsilon$ is the growth rate noise, with variance $\sigma_{\epsilon}^2$. The mean value of growth rate along lineages becomes $\langle \lambda\rangle =b/(1-a)$, with variance $\sigma_{\lambda}^2=\sigma_{\epsilon}^2/(1-a^2)$. The correlation coefficient between the mother-daughter growth rates is equal to $a$. It is clear that if $a=0$, we go back to the case without growth rate correlations, and as $a\rightarrow 1$, the growth rates become highly correlated.

Analytic computations are challenging due to the finite correlation of growth rates between generations. Thus, we first numerically confirm that our previous conclusions remain valid for correlated growth rates (non-zero $a$), namely, that the population growth rate is independent of the size control parameter $\alpha$ (Figure \ref{correlated_growth_rate}A), time-additive noise (Figure \ref{correlated_growth_rate}B) and size-additive noise (Figure S1). Our numerical simulations indicate that a positive correlation of growth rates tends to increase the population growth rate, as shown in Figure \ref{correlated_growth_rate}C. In the case where $a>0$, the fast growing cells tend to have more descendants -- which grow fast as well -- thus the mean growth rate of the leaf cells is increased in comparison with the case $a=0$. This enhances the population growth rate compared with the case of uncorrelated growth rates.

We find a threshold value of $a\approx 0.5$ separates two fundamentally different scenarios, one in which the variability in growth rates decreases the fitness (if $a$ is below 0.5), and another where it enhances it, see Figure \ref{correlated_growth_rate}C. Our general formula based on the tree distribution $f_0(\tau)$, Eq. (\ref{growthratef0}), is still valid, as expected. However, here even in the limit where growth rate fluctuations are the only source of noise, we do not have $f_0(\tau)=f(\tau)$, as in the case $a=0$, making it challenging to obtain a closed analytic expression for the population growth rate. Numerically, the resulting $\Lambda_p$ can be well fitted by the general formula,
\begin{equation}
\Lambda_p(\sigma_{\lambda})=\lambda_0\{1-C_1(a)\left(\frac{\sigma_{\lambda}}{\lambda_0}\right)^2\},\label{growthratea}
\end{equation}
where the single fitting parameter $C_1$ is shown in the inset of Figure \ref{correlated_growth_rate}C. Analyzing the experimental data of Ref. \cite{Stewart2005} on {\it E. coli} growth, we find a small Pearson correlation coefficient $a$, typically less than $0.1$, between mother and daughter growth rates (Figure S3). This places this biological scenario well within the weak correlation regime, and implies that Eq. (\ref{growthrate1}) for the uncorrelated case provides a good approximation for this case. Small correlation coefficients, well below $0.5$, are also reported in other works (\cite{Wang2010,Taheri2015}). Thus, it appears that for \emph{E. coli}, variability in the single-cell growth rates decreases the population fitness. Main theoretical and numerical results and notations are summarized in Table. \ref{table1} and Table. S1.

\section{Experimental test of the model on \emph{E. coli} data}
\begin{figure*}[bht]
   \center \includegraphics[width=1\textwidth]{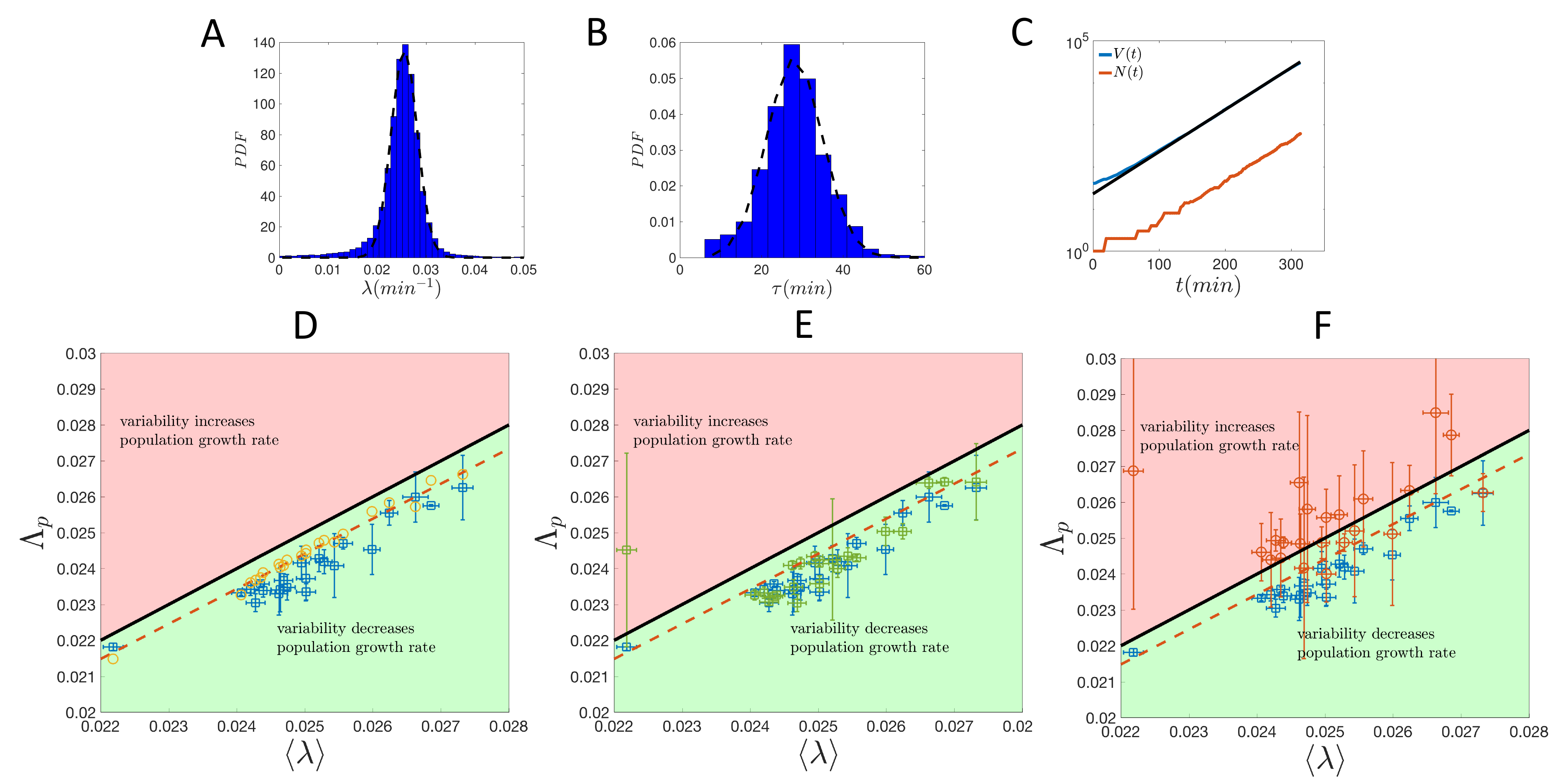}
\caption{{\bf Experimental tests of the model predictions using the data of Ref. \cite{Stewart2005}, courtesy of Eric Stewart.} Dimensions of both the \emph{x} and \emph{y} axes are $min^{-1}$ for D,E,F. \\
(A) The single-cell growth rate distribution has a CV about $0.19$, averaged over 22 samples with $N=635$ cells on average in each sample.\\
(B) The generation time distribution has CV of about $0.29$, averaged over samples.\\
(C) The time trajectories of total cell volumes ($V(t)$) and cell numbers ($N(t)$). The black line shows the fitted slope used to extract the population growth rate.\\
(D) Comparison between the theoretical prediction of $\Lambda_p$ (red dashed line), Eq. (\ref{growthrate1}) and the measured $\Lambda_p$ (blue square). The black line is the averaged single-cell growth rate, simply the line $y=x$. The yellow circles are the theoretical predictions from Eq. (\ref{growthratef0}) after replacing $\tau$ by $\ln(2)/\lambda$ using the experimentally inferred growth rate distribution with a bin size $10^{-3}$ $min^{-1}$. We used the results of a fit to data from the first half of the exponential phase to determine the error bars in the $y$ axis. The error bars in the $x$ axis is the standard deviation divided by the square root of sample size. \\
(E) The predictions from Eq. (\ref{growthratef0}) using the tree distribution directly (without replacing $\tau$ by $\ln(2)/\lambda$) are shown as green squares. The error bars of the green squares are computed using the differences obtained if the initial measurement time is modified to be $50$ mins instead of $150$ mins.\\
(F) The red circles are the predictions using the single lineage distribution of generation times, $f(\tau)$, which are typically far above the population growth rate. Cells after at least $5$ generations from the first ancestral cells are used to compute $f(\tau)$. The error bars of the red circles are computed using the differences obtained when the smallest generation difference of cells used in the analysis (from the first ancestral cell) is relaxed to $3$.}\label{experiments}
\end{figure*}

To test our theoretical and numerical predictions, we analyzed the data from Ref. \cite{Stewart2005} of {\it E. coli} growth on agarose pads. This is one of the most extensive datasets of \emph{E. coli} growth with complete lineage statistics. We extracted the distribution of single-cell growth rates and generation times along lineages, which for our purposes may be approximated by normal distributions: we will show that using the raw distributions of growth rates without assuming normality has little effect on the analysis. The experimental distributions and a Gaussian fit is shown in Figure \ref{experiments}A, B. The fits' coefficients of variation were $0.19$ and $0.29$, for the growth rates and generation times, respectively. { As mentioned previously, when working with actual data, it is more accurate to calculate the time-dependence of the measured total cell volume rather than cell number: the former varies continuously, while the latter shows strong transient and discrete effects. This is especially important for experiments in which the colony originates from a single-cell, in which cell division events remain approximately synchronized for many generations. For this reason we calculate the population growth rate from the growth of total cell volume.}
Our result is shown in Figure \ref{experiments}C, where as in Ref. \cite{Soifer2016} we used cells only in the steady-state regime, approximately reached at $t=150$ mins (see also Figure \ref{experiments}C). We plot the measured population growth rate against the mean single-cell growth rates for $22$ independent samples under the same experimental conditions, shown as the blue squares in Figure \ref{experiments}D. The red dashed line is the theoretical prediction of Eq. (\ref{growthrate1}), using the CV of the growth rate distribution obtained by pooling data from all experiments. Direct simulations of the size regulation model compared with theoretical predictions are shown in SI (STAR Methods, Figure S4). Note that the predicted relative difference between the population growth rate and averaged single-cell growth rate is only about $2\%$. Nevertheless, such an effect is crucial at the level of evolutionary dynamics, and would act to suppress growth-rate variability. Next, we also relax our assumption of a Gaussian growth rate distributions, from which Eq. (\ref{growthrate1}) is derived, and use the \emph{raw} growth rate distributions of each sample. We take Eq. (\ref{growthratef0}) and replace $\tau$ by $\ln(2)/\lambda$ to compute the population growth rate from the growth rate distribution, shown as the yellow circles in Figure \ref{experiments}D. We find little difference between the results using the Gaussian approximation for the growth rate distribution and those using the raw distribution of growth rates. Note that theoretically, we have shown in the main text that the generation time distribution itself, for a fixed growth rate distribution, has no effect on the population growth, hence we do not use the assumption of a normal generation time distribution in our predictions.
It seems plausible that the scatter of the experimental data around the theoretical prediction could partially come as an effect of the finite number of cells used to extract the population growth. To test this, we compared numerical simulations of our model with the analytic result, both for a relatively small number of cells comparable to that used in the experiments ($N\approx 600$) and a much larger number of cells. Figure S4 of the SI shows that indeed while the latter agrees well with our prediction, with no fitting parameters, the former shows a similar scatter around the theoretical results as that of the experimental data.
Finally, a main outcome of our analysis is that using statistics on single lineage is distinct from those on the entire tree. We sought to test this directly on the experimental dataset. In Figure \ref{experiments}E, we compute the theoretical prediction from Eq. (\ref{growthratef0}) using the tree distribution $f_0(\tau)$ (shown as the green squares), finding good agreement with the data. However, when using the lineage distribution $f(\tau)$, we obtain predictions which overestimate the measured population growth rate, as shown in Figure \ref{experiments}F. Taken together, these experiments support our central conclusions, and the relations we find between the growth rate distributions and the population growth.

\section{Discussion}
In this work, we investigated the effects of cell size regulation on the population growth. We extended the independent generation time (IGT) model to the realistic scenario where cell size is controlled. We considered primarily two sources of noises at the single-cell level: (i) growth rate fluctuations, (ii) time-additive or size-additive noises that affects the generation time while leaving the growth rate unchanged. We showed analytically, numerically as well as on experimental data, that a finite correlation between mother and daughter generation times exists, thus invalidating a central assumption of the independent generation time (IGT) model. We showed that as long as cells regulate their sizes, the population growth rate has to be equal to the total volume growth rate. We use this identity to prove that the population growth rate is equal to the single-cell growth rate in the case where growth rate fluctuations are negligible, in contrast to the prediction of the IGT model. In the presence of finite growth rate fluctuations, we find that the population growth rate is independent of the strength of the size control (governed by the regulation parameter $\alpha$), and the magnitudes of the time-additive or size-additive noises. $\Lambda_p$ thus only depends on the single-cell growth rate fluctuations, enabling us to calculate an explicit expression for it in the case where growth rates are uncorrelated across generations, which appears to be approximately realized in various experiments. Furthermore, we clarify the connections between the generation time distributions defined on lineages and trees, and are able to modify the results of the IGT model to account for the generic case -- including the case of correlated growth rates. Importantly, we predict that the variability in growth rates \emph{reduces} the population growth rate, as long as the correlation between the growth rates of mother and daughter cells is not too strong -- which appears to be a prevalent biological scenario.

Recent microfluidic experiments on \emph{E. coli} growth report a test of the IGT model, and the authors suggest that the population doubling time is modified in a manner consistent with the IGT model's prediction (\cite{Hashimoto2016}). Importantly, Hashimoto, {\it et al.} assumed the generation time of mother and daughter cells are independent. As we have shown, this assumption is inconsistent with the existence of size control because an independent fluctuating generation time will lead to divergent cell size. Furthermore, they do not make the distinction between the statistics on single lineages and trees, and consider them to be identical -- yet we have shown that making this distinction is crucial to obtain a correct interpretation of the results. Experimentally, the fact that they use all the data they record from the lineage trees instead of tracking a single lineage implies that they are approximately extracting the tree distribution, $f_0(\tau)$ (essentially using Eq. (\ref{mux}) in the STAR Methods). Their experimental results are therefore consistent with the relation Eq. (\ref{growthratef0}). While their experimental result and data analysis are consistent with our work, we provide what we believe to be the correct theoretical interpretation of these results. { We point out that the result of the IGT model that the population doubling time is smaller than the mean generation time is still intact. However, our work suggests that the relevant parameter to characterize the population growth is the single-cell growth rate rather the cell generation time. This is particularly clear in the limit of no growth rate fluctuations, in which the population growth rate precisely equals to the single-cell growth rate -- irrespective of the fluctuations in the generation time.}
%with an opposite conclusion: that the variability decreases rather than en- hances the population growth.}

 We independently tested our predictions on the experimental data of \emph{E. coli} growing on agarose pads, from Ref. \cite{Stewart2005}. We find that, indeed, the population growth (inferred from the exponential increase of the total cell volume) is \emph{smaller} than the average single-cell growth rate. Since the population growth rate is correlated with the population fitness, our results suggest that microbial populations would tend to {\it reduce} the growth rate variability given a constant mean growth rate, which may provide an explanation as to why the growth rate variability in nature is typically very small. In support of this, various works have reported very narrow growth rate distributions (\cite{Taheri2015,Cermak2016}), with the generation time distributions tending to be broader. { However, we cannot exclude that in other cases the growth rate variability may become beneficial due to a strong correlation between the mother and daughter growth rates.}

In the future, it would be intriguing to further explore the evolutionary consequences of our work, as well as put it in a physiological context -- what are the constraints on the variability and means of the growth rate distribution? {
In certain experiments the growth rate CV was reported independent of mean growth rate (\cite{Taheri2015, Biswas2014}), hence the relative population growth rate deficit is predicted to be independent of the growth rate, while in others (\cite{Kennard2016}) no such scaling was observed experimentally, hence the relative population growth deficit is expected to be a function of the growth rate itself.} Furthermore, the fitness reduction due to growth rate variability may also provide an additional constraint on the growth rate distributions used in genome-scale models of cellular metabolism (\cite{Martino2016}): { it would be beneficial to develop more mechanistic models which predict growth rate statistics (e.g., by describing metabolics and nutrient uptake explicitly) and where the tradeoff between mean and variability might emerge -- without having to dictate a constraint on mean growth rate explicitly. (in fact, experiments where growth rate variability can be controlled, without affecting the mean, would also be useful to this end).}
On the mathematical side, in this work we have extensively used the discrete Langevin approach (\cite{Amir2014, marantan2016stochastic}), yet other theoretical methods based on the so-called ``sloppy size control" have been proposed as well (\cite{Taheri2015, Kennard2016}), and shown to exhibit special scaling forms. Relations between the two formalisms have recently been worked out (\cite{Grilli2017}), and it would be interesting to explore whether further insight into these problems may be gained by using additional theoretical tools.
Finally, while in this work we considered \emph{intrinsic} variability, arising from the randomness associated with cellular processes, it will be interesting to investigate the effects of spatial or temporal environmental fluctuations. In such cases the growth rate variability can in fact be amplified, {e.g.}, by engineering an environment with spatial or (and) temporal variability in the nutrient availability.

\section{STAR METHODS}
Detailed methods are provided in the STAR Methods of this paper.
\section{AUTHOR CONTRIBUTIONS}
All authors conceived the work, carried out the work, and jointly wrote the manuscript.
\section{ACKNOWLEDGMENTS}
We thank Felix Barber, Po-Yi Ho, Jiseon Min, Sven van Teeffelen and Felix Wong for insightful discussions, and all reviewers for a careful reading of our manuscript and for their useful comments. We thank Eric Stewart for allowing us to use his data. AA thanks the A.P. Sloan foundation, the Milton Fund, the Volkswagen Foundation and Harvard Dean’s Competitive Fund for Promising Scholarship for their support.

\bibliographystyle{apalike}
\bibliography{lin}

%%%%%%%%%%%%%%%%%%%%%%%%%%%%%%%%%%%%%%%%%%%%%%

\clearpage
\onecolumngrid
\chapter{STAR METHODS}

Further information and requests for resources should be directed to and will be fulfilled by the Lead Contact, Ariel Amir (arielamir@seas.harvard.edu)
\section{Computing the correlation of mother-daughter cells' generation time}\label{method1}

It is convenient to replace Eq. (1) of the main text by the following relation (\cite{Amir2014}):
\begin{equation}
v_d=2\Delta ^{\alpha} v_b^{1-\alpha}. \label{approxeq}
\end{equation}
Both models agree to first order in a Taylor expansion with respect to the variable $v_b$, taken around the typical cell size at birth $\Delta$, for any value of $\alpha$.
Furthermore, the coefficient of variation (CV, the standard deviation divided by the mean) of cell birth sizes are often reported to be around $0.1$ (\cite{Campos2014}), indicating that the noise is relatively small and that the first order expansions makes for an excellent approximation, see the dashed line in Figure 1B of the main texts. Thus, the approximate model provides a very good realization of the original model. The corresponding generation time becomes
\begin{equation}
\tau=\frac{\ln 2}{\lambda}-\frac{\alpha}{\lambda}\ln \left(\frac{v_b}{\Delta}\right)+\xi\label{tauapprox}.
\end{equation}

 Next, we calculate the mother-daughter correlation, using the approximate model. Similar calculations have been done in Refs. \cite{Amir2014} and \cite{Taheri2015}. We consider the case without growth rate fluctuations. At generation $n+1$, $\tau(n+1)$ is determined by $v_b(n)$ and $\tau(n)$ as
\begin{equation}
\tau(n+1)=\frac{ln(2)}{\lambda_0}-\frac{\alpha}{\lambda_0}\ln\left(\frac{v_b(n)e^{\lambda_0\tau(n)}}{2\Delta}\right)+\xi_{n+1}.
\end{equation}
The auto-correlation between $\tau(n)$ and $\tau(n+1)$ becomes
\begin{equation}
\langle \tau(n+1)\tau(n)\rangle_c=-\frac{\alpha}{\lambda_0}\langle \ln\left(\frac{v_b(n)}{\Delta}\right) \tau(n)\rangle_c-\alpha\sigma_{\tau}^2,
\end{equation}
where $\langle AB\rangle_c=\langle AB\rangle-\langle A\rangle\langle B\rangle$, and $\sigma_{\tau}$ is the generation time standard deviation. Using Eq. (\ref{tauapprox}), we get $\langle \ln(v_b/\Delta)\tau\rangle_c=-\alpha\sigma_{v}^2/\lambda_0$, where $\sigma_{v}$ is the standard deviation of $\ln(v_b/\Delta)$, and obtain the correlation coefficient between the mother-daughter generation time as
\begin{equation}
C_{md}=\frac{\langle \tau(n+1)\tau(n)\rangle_c}{\sigma_{\tau}^2}=-\frac{\alpha}{2}.
\end{equation}

\section{A recursive derivation of the independent generation time model} Within the IGT model, each cell has a random generation time drawn from a given probability distribution $f(\tau)$. Imagine the population starts from a single cell at time $t=0$, and after some transient stage, the number of cells increases as $N(t)\sim \exp(\Lambda_p t)$. One can alternatively consider the population as two independent populations initiated by the first cell's two daughters. Assuming the first cell divides at time $\tau$, then the two sub-populations will increase with time as $N_1(t)\sim \exp(\Lambda_p(t-\tau))$. The two interpretations of the population must be equivalent after we average over all possible generation times, so $\exp(\Lambda_p t)=2\int_{0}^{\infty} f(\tau) \exp(\Lambda_p(t-\tau)) d\tau$, leading to the formula (\cite{Powell1956,Hashimoto2016,Biswas2016})
\begin{equation}
2\int_{0}^{\infty} f(\tau) e^{-\Lambda_p\tau} d\tau=1 .\label{IGTf}
\end{equation}
It is important to note that we have assumed that the number of cells in the subpopulations increases in the same manner as the original population, except for a temporal shift by the mother's generation time. This is only true for the IGT model. It can be proven that within the IGT model, the population doubling time is smaller than the mean generation time, namely, $\ln(2)/\Lambda_p \leq \langle \tau \rangle$, where $\langle \tau\rangle=\int_0^{\infty} \tau f(\tau)d\tau$ (\cite{Hashimoto2016}). The equality holds only when $f(\tau)=\delta(\tau-\langle\tau\rangle)$.

\section{Relations between different generation time distributions}

\subsection{Two kinds of cells: branch cell and leaf cell}
As shown in Fig. \ref{lineages_vs_trees}B in the main text, there are two different kinds of cells on the lineage tree, which are respectively the current generation (leaf cells) and cells from previous generations (branch cells). The following identity is very useful: $N_{leaf}=N_{branch}+1$, true for any lineage tree with binary division. Besides the generation time distribution along a lineage $f(\tau)$, we furthermore define $f_0(z)$, $f_1(z)$, $f_2(z)$ as the distributions of some cell cycle quantities $z$, respectively, for all cells in the tree (tree distribution), only branch cells (branch distribution), and only leaf cells (leaf distribution). Because for the whole populations $N_{leaf}\approx N_{branch}$, it is always true in the large $N$ limit that
\begin{equation}
f_0(z)=\frac{f_1(z)+f_2(z)}{2}.
\end{equation}

%\begin{figure}[hbt!]
%   \centering \includegraphics[width=0.4\textwidth]{growthrate_on_nl_different_ns_alpha_05_oct_5}
%\caption{$\Lambda_p$ \it {vs.} $\sigma_{\lambda}$, with two different size-additive noises $\sigma_{\eta}$. Here, we fix the time-additive noi1se as $\sigma_{\xi}=0.05$.}\label{growthrate1}
%\end{figure}

\subsection{Age distribution: $\phi(x)$}
In the exponential growth phase of the population, each cell essentially plays the same role and can be considered as the leading cell of the following lineage tree. One can therefore use the tree distribution $f_0(\tau)$ as the generation time distribution of each cell on the tree, and if there is no mother-daughter correlation, $f_0(\tau)$ is equal to $f(\tau)$. In the following, we essentially follow the same idea originally by Powell (\cite{Powell1956,Hashimoto2016}), with the key difference of replacing the lineage distribution $f(\tau)$ by the tree distribution $f_0(\tau)$.

Given $f_0(\tau)$, we define the survival function, which gives the probability for cells to survive at least to the age $x$
\begin{equation}
F_{-}(x)=\int_{x}^{\infty} f_0(y) dy,
\end{equation}
and the division rate as
\begin{equation}
\mu(x)=\frac{F_{-}(x)-F_{-}(x+dx)}{F_{-}(x) dx}=\frac{f_0(x)}{F_{-}(x)}.\label{mux}
\end{equation}
We can alternatively express $F_-(x)$ as $F_-(x)=\exp(-\int_0^{x}\mu(y)dy)$.

For large times, the age distribution $\phi(x,t)$ will approach a stationary distribution $\phi(x)$. The number of cells at time $t$ at age $x$ is thus $N(t)\phi(x)$, and the probability for them to live to $t+dt$ is $1-\mu(x)dt$. In the exponential phase, the number of cells increases exponentially as $N(t)\sim \exp(\Lambda_p t)$, with the population growth rate $\Lambda_p$. In this way we can find the self-consistent equation for the age distribution
\begin{equation}
\frac{N(t)\phi(x)(1-\mu(x)dt)}{N(t+dt)}=\phi(x+dt) ,
\end{equation}
leading to the equation of $\phi(x)$ as
\begin{equation}
\frac{d\phi(x)}{dx}=-\mu(x) \phi(x)-\Lambda_p \phi(x).
\end{equation}
Using $F_-(x)=\exp(-\int_0^{x}\mu(y) dy)$, we find the solution as $\phi(x)=\phi_0 \exp(-\Lambda_p x) F_{-}(x)$ with $\phi_0$ to be determined. The population growth rate can be expressed as the integral of age distribution and division rate
\begin{align}
\Lambda_p&=\frac{N(t)\int_{0}^{\infty} \phi(x)\mu(x) dx}{N(t)}  \nonumber \\
&=\int_{0}^{\infty} \phi(x)\frac{f_0(x)}{F_{-}(x)} dx=\int_{0}^{\infty} \phi_0 e^{-\Lambda_p x} f_0(x) dx,\label{lambdap2}
\end{align}
because it is the cells' divisions that increase the number of cells.

From the normalization of $\phi(x)$, we get
\begin{align}
1=&-\frac{\phi_0}{\Lambda_p}\{-1+\int_{0}^{\infty} e^{-\Lambda_p x} f_0(x) dx\}\nonumber \\
=&\frac{\phi_0}{\Lambda_p}-\frac{\phi_0}{\Lambda_p}\int_{0}^{\infty}e^{-\Lambda_p x}f_0(x)dx.
\end{align}
Combining with Eq. (\ref{lambdap2}), we get $\phi_0=2\Lambda_p$, and
\begin{equation}
\phi(x)=2\Lambda_p e^{-\Lambda_p x} F_{-}(x)
\end{equation}

\subsection{Ancestors' generation time distribution: $f_1(\tau)$}
In this subsection, we derive the expression of $f_1(\tau)$, given tree distribution $f_0(x)$. Imagine we review a snapshot at some time $t^{\prime}$ before the current time $t$, and all the cells at time $t^{\prime}$ are branch cells. Let's first write down the expression of $f_1$ and explain its interpretation,
\begin{equation}
f_1(\tau)=\frac{N(t^{\prime}) \phi(\tau) \mu(\tau) d\tau}{N(t^{\prime})\Lambda_p d\tau} \label{f11}
\end{equation}

Here, the denominator represents the number of division events from $t^{\prime}$ to $t^{\prime}+d\tau$. The numerator represents cells which divide at age $\tau$ at time $t^{\prime}$. Their ratio becomes the fraction of branch cells that divide at age $\tau$, which means that they have a generation time equal to $\tau$. The above derivation does not apply to leaf cells, since they have not divided yet. One should note that here we are not making any assumptions about the correlation between mother-daughter generation time since we focus on one single generation, so Eq. (\ref{f11}) is a universal result. Using the expression of $\phi(x)$, we obtain
\begin{equation}
f_1(\tau)=2e^{-\Lambda_p \tau}f_0(\tau),\label{f1_sp}
\end{equation}
and we get the equation to extract the population growth rate $\Lambda_p$ from the normalization of $f_1(\tau)$,
\begin{equation}
\int_{0}^{\infty} 2e^{-\Lambda_p \tau}f_0(\tau) d\tau =1,\label{lambdap1}
\end{equation}
as verified in the main text.

\subsection{Current cells' generation time distribution: $f_2(\tau)$}
We first write down the expression of $f_2(\tau)$
\begin{equation}
f_2(\tau)=\frac{\int_{0}^{\tau} N(t)\phi(y) \frac{F_-(\tau)}{F_-(y)}\mu(\tau)dy}{N(t)},
\end{equation}
here in the numerator, the expression inside the integral means those cells which are at age $y$ at time $t$ and will survive until age $\tau$ and divide. Given $\phi(\tau)$, it is easy to obtain
\begin{equation}
f_2(\tau)=2f_0(\tau)(1-e^{-\Lambda_p \tau}).\label{f2_sp}
\end{equation}
Again here we do not make any assumption about the mother-daughter correlation. From Eqs. (\ref{f1_sp}, \ref{f2_sp}), we indeed find that
\begin{equation}
f_0(\tau)=\frac{f_1(\tau)+f_2(\tau)}{2}.	
\end{equation}

\section{Computing the population growth rate}
 In the limit $\sigma_{\xi}\rightarrow0$, and a small but finite $\sigma_{\lambda}\ll \lambda_0$, we are able to calculate the analytic expression of the population growth rate $\Lambda_p$. Because the correlation between mother-daughter cells' generation time  is zero, we can use the IGT model's result, Eq. (\ref{IGTf}). In this case, $\tau=\ln(2)/\lambda$ because the birth size $v_b$ converges to $2\Delta$ very quickly without time-additive or size-additive noise, and there are essentially no size fluctuations. We can redefine the variable $x=\ln(2)/\tau$, and the formula to calculate $\Lambda_p$ becomes
\begin{equation}
2\int_0^{\infty} \frac{1}{\sqrt{2\pi\sigma_{\lambda}^2}}\exp(-\frac{\ln(2)\Lambda_p}{x})\exp(-\frac{(x-\lambda_0)^2}{2\sigma_{\lambda}^2})dx=1.
\end{equation}

We can rewrite the integral without the prefactor as
\begin{align}
I&=\int_0^{\infty} \exp(-\frac{1}{2\sigma_{\lambda}^2}((x-\lambda_0)^2+\frac{2\sigma_{\lambda}^2\Lambda_p\ln(2)}{x}))dx\nonumber\\
&=\int_0^{\infty} \exp(-\frac{1}{2\sigma_{\lambda}^2}g(x))dx,
\end{align}
and use the saddle point method to calculate the integral,
\begin{equation}
I=\exp(-\frac{1}{2\sigma_{\lambda}^2} g(x_c)) \sqrt{\frac{4\pi\sigma_{\lambda}^2}{g^{\prime\prime}(x_c)}}.\label{saddle}
\end{equation}
$x_c$ is determined by $g^{\prime}(x_c)=0$, from which we get $x_c\approx \lambda_0+\frac{\sigma_{\lambda}^2\Lambda_p\ln(2)}{\lambda_0^2}$ and
\begin{align}
g(x_c)&\approx \frac{2\sigma_{\lambda}^2\Lambda_p\ln(2)}{\lambda_0+\frac{\sigma_{\lambda}^2\Lambda_p\ln(2)}{\lambda_0^2}}+\frac{\sigma_{\lambda}^4\Lambda_p^2\ln(2)^2}{\lambda_0^4},\\
g^{\prime\prime}(x_c)&\approx 2+\frac{4\sigma_{\lambda}^2\Lambda_p\ln(2)}{(\lambda_0+\frac{\sigma_{\lambda}^2\Lambda_p\ln(2)}{\lambda_0^2})^3}.
\end{align}
Keeping the lowest correction due to $\sigma_{\lambda}$, we eventually find
%\begin{equation}
%1-\frac{\Lambda_p}{\lambda_0}-\frac{\sigma_{\lambda}^2\Lambda_p}{\lambda_0^3}+\frac{\sigma_{\lambda}^2\Lambda_p^2\ln(2)}{2\lambda_0^4}=0
%\end{equation}
%Again, we only keep the lowest correction due to $\sigma_{\lambda}$ to determine $\Lambda_p$,
\begin{equation}
\Lambda_p=\lambda_0\{1-(1-\frac{\ln(2)}{2})\left(\frac{\sigma_{\lambda}}{\lambda_0}\right)^2\}.\label{lambdap}
\end{equation}

\section{Robustness of the results to other form of noise distributions}
\subsection{Effects of size-additive noises}\label{sizenoise}
In the main text, we considered the effects of time-additive noise and single-cell growth rate variability. We may also investigate the effects of \emph{size-additive} noise, by using the following stochastic equation:
\begin{equation}
v_d=2\alpha \Delta +2(1-\alpha)v_b + 2\eta.\label{sizenoise1}
\end{equation}
Here, $\eta$ is the size-additive noise, with zero mean and variance $\sigma_{\eta}^2$. We fix the time-additive noise as $\sigma_{\xi}=0.05$, and change $\sigma_{\eta}$, $\sigma_{\lambda}$. Numerical simulations support that the size-additive noise does not affect the population growth rate (Figure S1A,B). This implies that we may approximate the population growth rate using Eq. (\ref{growthrate1}) of the main text. Moreover, we find that the generation time distribution in the presence of size-additive noise exhibits an exponential tail for large generation times (Figure S2). This is similar to the prediction of Ref. \cite{Pugatch2015}.

\subsection{Effects of non-Gaussian time-additive noises}\label{nongaussiantime}
Furthermore, we may also consider a Gamma-distributed time-additive noise (\cite{Hashimoto2016}), so
\begin{equation}
\tau=\frac{1}{\lambda}\ln \left( \frac{v_d}{v_b}\right) +\xi ,
\end{equation}
where $\xi$ is a random number following the distribution,
\begin{equation}
\rho(\xi)=\frac{1}{b^a\Gamma(a)}(\xi+ab)^{a-1}e^{-\frac{\xi+ab}{b}}.
\end{equation}
Here, we take $a=1$ and $b=\sigma_{\xi}/\sqrt{a}$ where $\sigma_{\xi}^2$ is the variance of the time-additive noise. The resulting generation time distribution is shown in Figure S1C. Even though the generation time distribution can be highly skewed, the population growth rate $\Lambda_p$ remains invariant (Figure S1D).

\subsection{Effects of non-Gaussian distributed growth rates}\label{nongaussianlambda}
 We also consider log-normal distributed single-cell growth rates,
\begin{equation}
\rho(\lambda)=\frac{1}{\sqrt{2\pi \sigma_{\lambda}^2}\lambda}\exp(-\frac{\ln(\lambda)^2}{2\sigma_{\lambda}^2}).
\end{equation}
As before, we can compute the population growth rate by setting $\tau=\ln(2)/\lambda$ in $2\int_{0}^{\infty}f_0(\tau)\exp(-\Lambda_p\tau)=1$, namely $2\int_{0}^{\infty}\rho(\lambda)\exp(-\ln(2)\Lambda_p/\lambda)=1$. Using similar saddle point approximations as the normal case and replace $y=\ln(\lambda/\lambda_0)$, we obtain the essentially the same result as Eq. (\ref{growthrate1}),
\begin{equation}
\Lambda_p=\langle\lambda\rangle \{1-(1-\frac{\ln(2)}{2})\left(\frac{\sigma_{\lambda}}{\lambda_0}\right)^2\},\label{lognormal}
\end{equation}
where the mean growth rate $\langle \lambda\rangle=\lambda_0\exp(\sigma_{\lambda}^2/2)$. The numerical results and theoretical predictions are shown in Figure S1E.

\section{Simulations \emph{vs.} Experiments}\label{simulation}
We simulated the size regulation model directly and chose $\sigma_{\lambda}=0.19$, $\sigma_{\xi}=0.12$ to have the same CV of growth rates and generation times as the experimental data (\cite{Stewart2005}) (as in Figure \ref{sizeregulation}A of the main text). We first simulate a small sample starting from a single cell and chose the same number of cells as in the experiments ($N\approx 600$), as shown in Figure S4A. The population growth rate is computed after $t=5$. The error bars of the simulations are inferred from a comparison with $\Lambda_p$ computed after $t=3$. Next, we simulated a larger sample starting from $1000$ cells and with $N\approx 2.6\times 10^{6}$ at the end of the simulation, as shown in Figure S4B.

\newpage
\section{Supplementary Informaton}

\renewcommand{\theequation}{S\arabic{equation}}
\setcounter{equation}{0}

\renewcommand{\thefigure}{S\arabic{figure}}
\setcounter{figure}{0}

\renewcommand{\thetable}{S\arabic{table}}
\setcounter{table}{0}

%\begin{figure}[hbt!]
%   \centering \includegraphics[width=0.7\textwidth]{lambda_p_on_sigmal_change_a_sep_28}
%\caption{Related to Figure 2B. $\Lambda_p$ \it {vs.} $\sigma_{\lambda}$, with fixed time-additive noise $\sigma_{\xi}=0.15$, and zero size-additive noise. The mean value of single-cell growth rate is fixed to be $1$, so $b=1-a$. $\sigma_{\epsilon}$ is adjusted according to $\sigma_{\lambda}$. The parameter $a$ introduced in Eq. (\ref{a}), changes from $0$ to $0.9$ with step $0.1$ as indicated in the figure, and $a=0.5$ is shown in filled square.}\label{growthrate2}
%\end{figure}

\begin{figure*}[bht!]
\center   \includegraphics[width=1\textwidth]{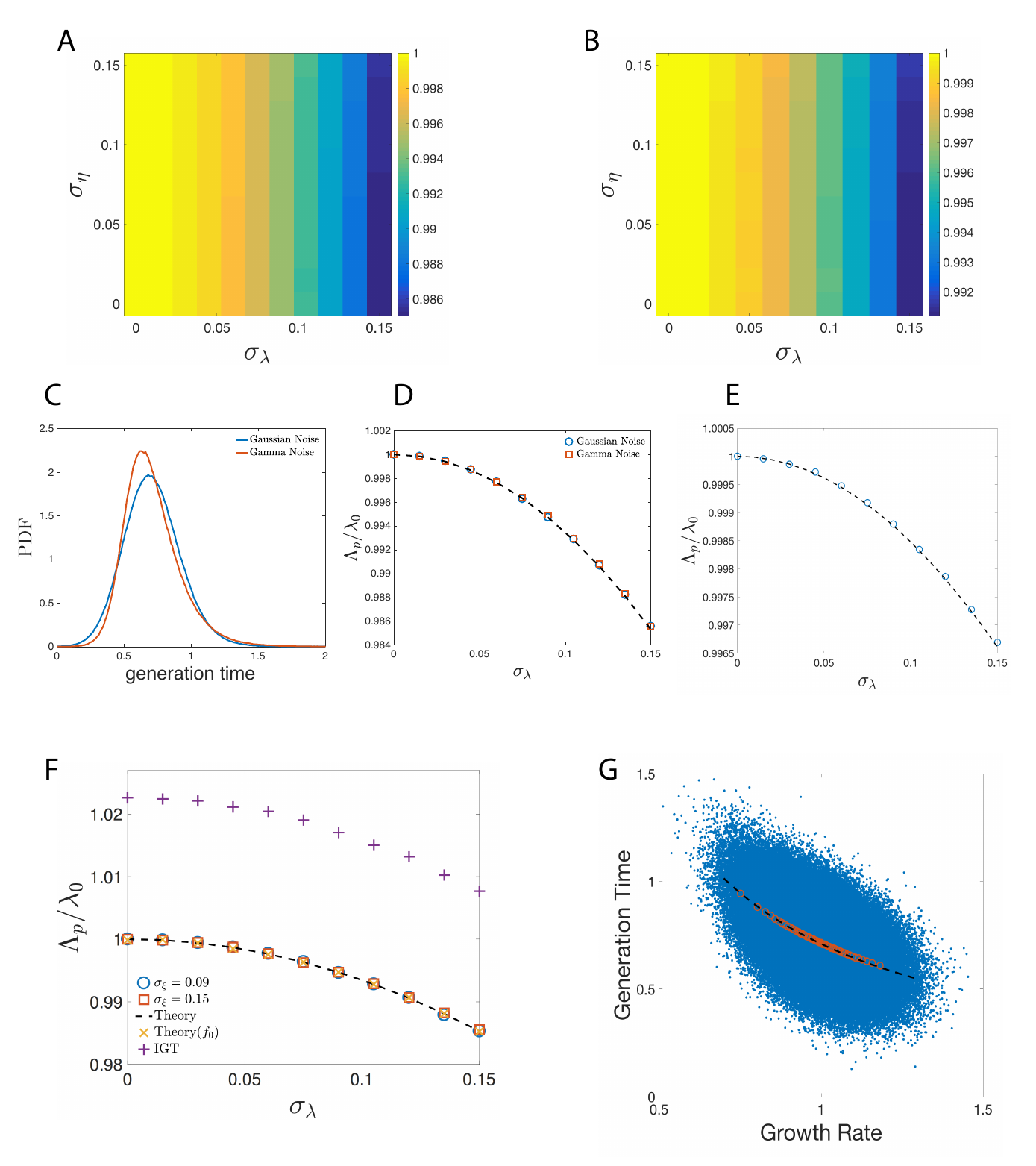}
\caption{Related to Figure 3. \\
{\bf (A, B)} The dependence of the population growth rate $\Lambda_p/\lambda_0$ on the growth rate variability $\sigma_{\lambda}$ and the size-additive noise $\sigma_{\eta}$. A: $a=0$ (no correlation between the growth rates of mother and daughter cells). B:$a=0.2$. $\sigma_{\xi}=0.05$ for both. \\
{\bf (C,D)} Effects of non-Gaussian distributed time-additive noises.
(C) The generation time distribution with Gaussian and Gamma distributed noises with the same variances.
(D) The resulting population growth rate is invariant to the form of noise distribution.\\
{\bf (E)} Population growth rate $\Lambda_p$  \emph{vs.} the growth rate variability $\sigma_{\lambda}$. The distribution of growth rates is log-normal and the solid line is the theoretical prediction, using Eq. (44) of the STAR Methods.\\
{\bf (F) } Predictions of IGT model ($\sigma_{\xi}=0.15$) is compared with the one from the tree distribution. \\
{\bf (G) }Single-cell growth rates \emph {vs.} the generation times of a population. The simulation starts from $1000$ cells and stops at $t=6$ with $N \approx 4\times 10^{5}$ cells. The data points are from the leaf cells, namely the instantaneous population. The Pearson correlation coefficient is about $-0.53$. The red circles are the binned data and the dashed line is an empirical fit, $\tau\approx 0.71/\lambda \approx \ln(2)/\lambda$. $\sigma_{\xi}=0.1$, $\sigma_{\lambda}=0.1$. \\}\label{sp1}
\end{figure*}

\begin{figure*}[bht!]
\center   \includegraphics[width=0.6\textwidth]{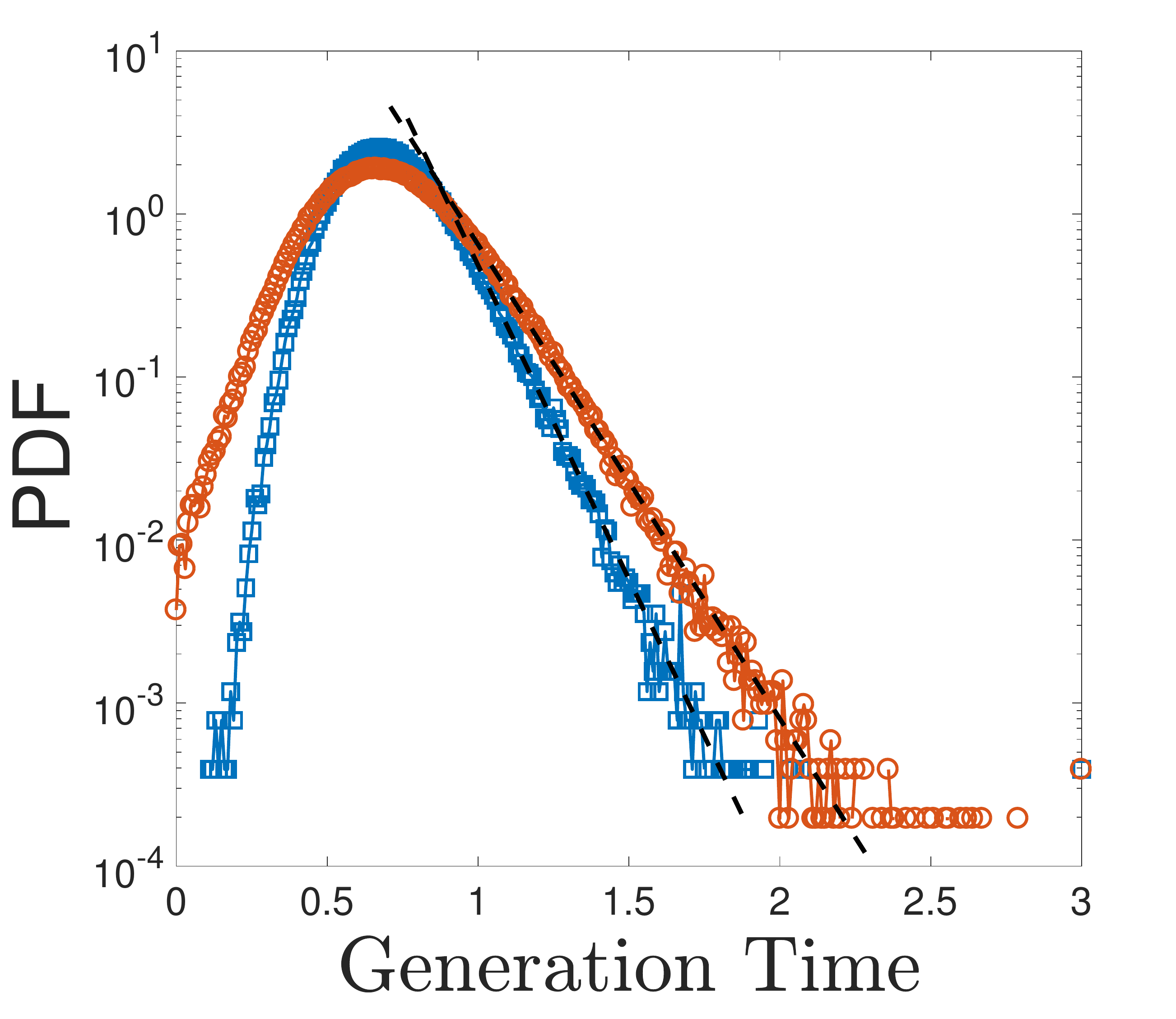}
\caption{Related to Figure 2. Generation time distributions in the presence of size-additive noise. The dashed lines indicate the exponential tails. Blue squares ($\sigma_{\eta}=0.09$). Red circles ($\sigma_{\eta}=0.15$). $\sigma_{\xi}=0.05$ and $\sigma_{\lambda}=0.15$ for both.}\label{sp2}
\end{figure*}

%\begin{figure*}[bht!]
% \includegraphics[width=0.9\textwidth]{figure_sp_3.pdf}
%\center \caption{{\blue Related to Fig. 3B. Effects of non-Gaussian distributed time-additive noises (Section \ref{nongaussiantime}).
%(A) The generation time distribution with Gaussian and Gamma distributed noises with the same variances.
%(B) The resulting population growth rate is invariant to the form of noise distribution.}}\label{sp3}
%\end{figure*}

%\begin{figure*}[bht!]
%\center   \includegraphics[width=0.6\textwidth]{figure_sp_4}
%\caption{{\blue Related to Fig. 3B. Population growth rate $\Lambda_p$  \emph{vs.} the growth rate variability $\sigma_{\lambda}$ (Section \ref{nongaussianlambda}). The distribution of growth rates is log-normal and the solid line is the theoretical prediction, using Eq. (11) of the main text.}}\label{sp4}
%\end{figure*}

\begin{figure*}[bht!]
\center   \includegraphics[width=0.6\textwidth]{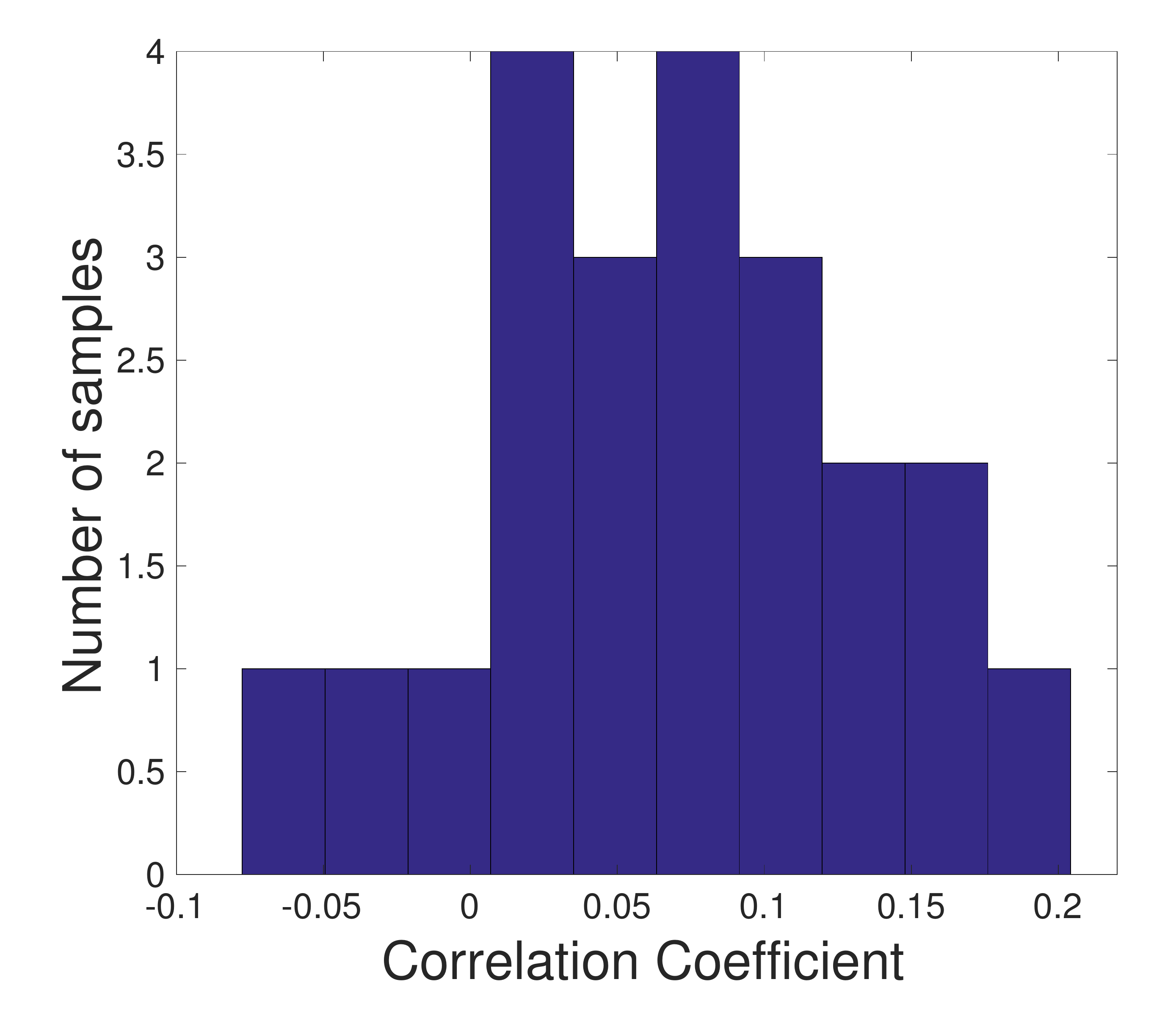}
\caption{Related to Figure 4. Histogram of the correlation coefficients of the single-cell growth rates between mother and daughter cells (data from Stewart et al., 2005). The histogram is based on $22$ runs of experiments. The mean value of the coefficient is $0.07$ with the standard deviation $0.07$.}\label{sp5}
\end{figure*}

%\begin{figure*}[bht!]
%\center   \includegraphics[width=0.6\textwidth]{figure_sp_6}
%\caption{{\blue Related to Fig. 3B. Predictions of IGT model ($\sigma_{\xi}=0.15$) is compared with the one from the tree distribution (Section \ref{IGTprediction}). }}\label{sp6}
%\end{figure*}

%\begin{figure*}[bht!]
%\center   \includegraphics[width=0.6\textwidth]{figure_sp_7}
%\caption{{\blue Related to Fig. 3B. Single-cell growth rates \emph {vs.} the generation times of a population (Section \ref{correlationtimelambda}). The simulation starts from $1000$ cells and stops at $t=6$ with $N \approx 4\times 10^{5}$ cells. The data points are from the leaf cells, namely the instantaneous population. The Pearson correlation coefficient is about $-0.53$. The red circles are the binned data and the dashed line is an empirical fit, $\tau\approx 0.71/\lambda \approx \ln(2)/\lambda$. $\sigma_{\xi}=0.1$, $\sigma_{\lambda}=0.1$. }}\label{sp7}
%\end{figure*}

\begin{figure*}[bht!]
\center   \includegraphics[width=1\textwidth]{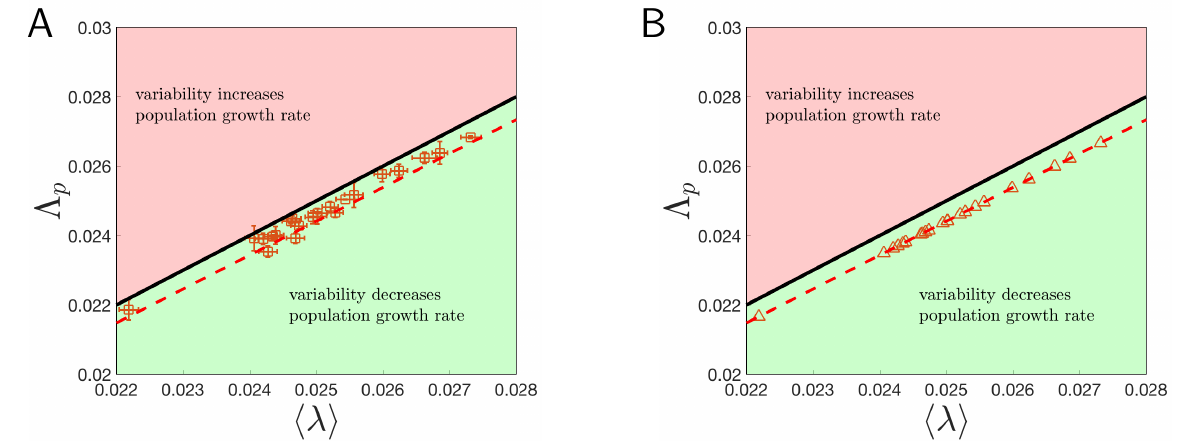}
\caption{Related to Figure 5. Simulation results of the size regulation model with $\sigma_{\lambda}=0.19$ and $\sigma_{\xi}=0.12$. (A) Simulations starting from one single cell and ending with the same number of cells as in the experiments of Stewart et al., 2005. Red squares are simulation results and the red dashed line is the theoretical prediction,  Eq. (10) in the main text. The error bars in y axis of the simulations are inferred from a comparison with the measured population growth rate computed after $t = 3$. The error bars in the x axis is the standard error (standard deviation divided by the square root of sample size). (B) Simulations starting from $1000$ cells and ending with $2.6\times 10^{6}$ cells. Red triangles are simulation results.}\label{sp8}
\end{figure*}
\begin{table*}
\caption{Summary of the main parameters of the size regulation model. Related to Table 1.}
\center\begin{tabular}{ l  | c |  c   }
\hline
                &Name  & Equation\\ [3pt] \hline
  $\alpha$&  Size control parameter  & $v_d=2\alpha \Delta+2(1-\alpha)v_b$ (Eq. (1))\\ [3pt] \hline
  $\Delta$ & (approx.) Mean cell volume at birth  & $v_d=2\alpha \Delta+2(1-\alpha)v_b$ (Eq. (1))\\ [3pt] \hline
  $\sigma_{\xi}$ & Standard deviation of time-additive noise & $\tau=\frac{1}{\lambda}\ln(\frac{v_d}{v_b})+\xi$ (Eq. (2))\\ [3pt] \hline
 $\sigma_{\lambda}$ & Standard deviation of growth rate noise & $\tau=\frac{1}{\lambda}\ln(\frac{v_d}{v_b})+\xi$ (Eq. (2)) \\ [3pt] \hline
  $\sigma_{\eta}$ & Standard deviation of size-additive noise & $v_d=2\alpha \Delta +2(1-\alpha)v_b + 2\eta$ (Eq. (40)) \\ [3pt] \hline
   $a$ & Correlation coefficient between growth rates & $\lambda_d=a\lambda_m+b+\epsilon$ (Eq. (12))  \\ [3pt] \hline
\end{tabular}\label{table1}
\end{table*}

\end{document}